\documentclass{article}
\usepackage[utf8]{inputenc}
\usepackage{mathtools}
\usepackage{amsfonts} 
\usepackage{amssymb}
\usepackage{appendix}
\usepackage{amsmath}
\usepackage[pdftex]{graphicx}
\usepackage{tikz-cd}
\usepackage{mathrsfs}
\usepackage{titlesec}
 \usepackage{pstricks-add}
 \usepackage{comment}
 \usepackage{epsfig}
 \usepackage[nottoc,notlof,notlot]{tocbibind} 
\usepackage{pst-grad} 
\usepackage{pst-plot} 
\usepackage[space]{grffile} 
\usepackage{etoolbox} 
\usepackage{hyperref}
\usepackage{color}
\usepackage{amsthm}
\hypersetup{
    colorlinks,
    citecolor=black,
    filecolor=black,
    linkcolor=black,
    urlcolor=black
}
\usepackage{cleveref}
\crefrangelabelformat{equation}{(#3#1#4--#5#2#6)}
\crefname{equation}{}{}
\Crefname{equation}{Equation}{Equations}
\newcommand{\der}[2]{\frac{\mathrm{d}}{\mathrm{d#1}}#2}
\newcommand{\derk}[3]{\frac{\mathrm{d}^{#3}}{\mathrm{d#1}^{#3}}#2}
\def\R{\mathbb{R}}
\def\pman{(M, \{\cdot,\cdot\})}

\def\Bvec{\mathbf{B}}
\def\Jvec{\mathbf{J}}

\def\inv{^{-1}}

\def\qvec{\mathbf{q}}
\def\pvec{\mathbf{p}}
\def\lvec{\mathbf{\Lambda}}
\def\Jvec{\mathbf{J}}
\def\Avec{\mathbf{A}}

\def\Qvec{\mathbf{Q}}
\def\Pvec{\mathbf{P}}

\def\uvec{\mathbf{u}}
\def\vvec{\mathbf{v}}
\def\Cin{C^{\infty}}

\def\kepphase{T^*(\mathbf{R}^3\setminus\{0\})}

\def\del{\circ d}

\def\lco{L\'{a}zaro-Cam\'{\i}\:and\:Ortega\:}
\def\Zvec{\mathbf{Z}}

\def\Xvec{\textbf{X}}
\def\TxS{T^\times S^3_{np}}
\def\ToneS{T^{\times}_1S^3_{np}}

\begin{document}
\title{Moser Regularization of a Stochastically Perturbed Kepler Problem}
\author{Archishman Saha}
\date{}
\newtheorem{thm}{Theorem}
\newtheorem{cor}{Corollary}
\newtheorem{lem}{Lemma}
\newtheorem{defn}{Definition}
\newtheorem{example}{Example}
\newtheorem{prop}{Proposition}
\newtheorem{rem}{Remark}
\newtheorem{theorem}{Theorem}
\newtheorem{remark}{Remark}
\newtheorem{proposition}{Proposition}
\newtheorem{lemma}{Lemma}
\newtheorem{corollary}{Corollary}
\newtheorem{definition}{Definition}
\newtheorem{conjecture}{Conjecture}
\maketitle
\abstract{We consider a stochastic Kepler problem perturbed by a
Hamiltonian noise affecting the angular momentum vector. We show that the angular momentum and the Laplace-Runge-Lenz vectors are
conserved in magnitude and as a consequence, the distance and speed of the particle follow deterministic dynamics. Further, in a procedure similar to Moser's regularization,
we transform the stochastic Kepler problem to obtain its dynamics as a
stochastic geodesic flow on a 3-sphere.}



\maketitle

\section{Introduction}
 The literature on stochastic mechanics may be broadly classified into two points of view: the first involves a kinematical description of stochastic processes and the second involves studying mechanical systems perturbed by random noise. The former involves the use of forward and backward derivatives to provide a dynamical description of diffusion processes, much akin to Schr\"{o}dinger's equation in quantum mechanics. This approach has been explored by Nelson \cite{nelson, nelson2}, Yasue \cite{yasue}, Arnaudon and Cruzeiro \cite{anabela}, Arnaudon et. al. \cite{arn}, Zambrini \cite{zambrini1986variational}, Huang and Zambrini \cite{huangzambrini} and others.
 
\medskip

The present study follows the latter point of view. Mechanical systems perturbed by external noise were studied from the Hamiltonian perspective by Bismut \cite{bismut} on Euclidean spaces and more generally, on Poisson manifolds by \lco \cite{lco1, lco2}. The variational principle side of perturbative stochastic mechanics was explored in the works of Bou-Rabee and Owhadi \cite{Bou_Rabee_2008}, Holm \cite{holm1}, Arnaudon et. al. \cite{holm2}, Cruzeiro et. al. \cite{holm3}, Crisan and Street \cite{crisan} and Street and Takao \cite{street2023semimartingale}.

\medskip

This work considers a stochastically perturbed Kepler problem. Stochastic versions of the Kepler problem has been studied both from an intrinsic as well as a perturbative point of view by Albeverio et. al. \cite{albeverio, Albeverio140040}, Nottale \cite{nottale1995new}, Nottale et. al. \cite{nottale1997scale}, Cresson \cite{cresson2011stochastisation}, Cresson and S{\'e}bastien \cite{cresson2007stochastic}, Cresson et. al \cite{cresson2015sharma}, Cresson, Nottale and Lehner \cite{cresson2021stochastic}.

\medskip
The problem we study concerns the Kepler problem perturbed by a noise affecting the angular momentum vector only. 
Specifically,  the Hamiltonian vector field of the Kepler problem is perturbed by noisy Hamiltonian vector fields arising out of the components of the angular momentum vector.

\medskip
The interesting fact about such a perturbation is that, even though the symmetries of the Kepler problem (the angular momentum and the Laplace-Runge-Lenz vector) are no longer conserved, the distance of the particle from the gravitational source as well as its speed behaves deterministically.  Another feature is that 
the momentum vector is always at a constant distance away from a stochastic process on a sphere, providing an analogue of the deterministic Hamilton's theorem (see Milnor \cite{milnor}).

\medskip
Given the deterministic behaviour of the distance and speed, collisions are well-defined. Recall that in case of the deterministic Kepler problem, Moser regularization transforms Keplerian orbits at a fixed negative energy level to orbits of the geodesic flow on the 3-sphere(see Moser \cite{Mo70}, Cushman and Bates \cite{cb} and Heckmann and de Laat \cite{heckdelaat}).
In our problem,  we are able  to carry out a transformation similar to Moser's, whereby orbits for a fixed negative energy level are transformed to the geodesic flow on the 3-sphere. Moreover, the transformation is ``noise-preserving" in the sense that the structure of the noise perturbation remains unchanged after applying the Moser map.

\medskip

The plan of the paper is as follows: In Section \ref{stochastichamiltonian}, we discuss stochastic perturbations of Hamiltonian systems as well as conserved quantities. The next two sections deal with the stochastic Kepler problem, Hamilton's theorem, collisions and the Moser transformation and regularization.

\section{Stochastic Perturbations of Hamiltonian Systems}\label{stochastichamiltonian}
Let $(\mathcal{S}, \mathcal{F}, \mathcal{P})$ denote a probability space. Let $\pman$ be a Poisson manifold. Assume that $h:M\rightarrow\R$ is a Hamiltonian. We want to perturb the Hamiltonian vector field $X_{h}$ by ``noisy'' Hamiltonian vector fields. For $i = 1, \cdots, k$, let $h_i\in\Cin(M)$. We consider the following SDE:

\begin{equation}\label{pertham}
    \del \Gamma_t = X_{h}(\Gamma_t) dt + \sum_{i = 1}^k X_{h_i}(\Gamma_t) \del B_t^i.
\end{equation}
where $B^i_t$ is a Brownian motion and $\del$ denotes Stratonovich integration. The reader is referred to Emery \cite{emerybook} for an introduction to  on manifolds, and Lázaro-Camí and Ortega \cite{lco2} for a treatment of the general theory of stochastic Hamiltonian systems.

\medskip
Given $f\in\Cin(M)$ and a solution $\Gamma_t$ of \eqref{pertham}, the evolution of $f$ along the solution $\Gamma_t$ can be computed by using Stratonovich or Itô integrals. 
\begin{proposition}
Assume the setup of equation \eqref{pertham}. Suppose $\Gamma_t$ solves the SDE \eqref{pertham} with initial condition $\Gamma_0$. Then, for every $f\in\Cin(M)$,
        \begin{equation}\label{stratrepoffbrown}
            f(\Gamma_t) - f(\Gamma_0) = \int_0^t\{f,h\}(\Gamma_s)ds+\sum_{i = 1}^{k}\int_0^t \{f, h_i\} (\Gamma_s)\del B^i_s,
        \end{equation}
        where the integral on the right is the Stratonovich integral and
        \begin{align}\label{itorepoffbrown}
            f(\Gamma_t) - f(\Gamma_0) = \int_0^t\{f,h\}(\Gamma_s)ds&+\sum_{i = 1}^{k}\int_0^t \{f, h_i\}(\Gamma_s) d B^i_s\\ \nonumber &+ \frac{1}{2}\sum_{i = 1}^k\int_0^t\{\{f, h_i\}, h_i\}(\Gamma_s)ds,
        \end{align}
        where the integral on the right is the Itô integral.
\end{proposition}
\noindent This follows from Proposition 2.2 and Proposition 2.3 in Lázaro-Camí and Ortega \cite{lco1}.

\medskip
A \textbf{conserved quantity} of the Stratonovich SDE \eqref{pertham} is a function $f\in\Cin(M)$ such that $f(\Gamma_t)$ is constant for any solution $\Gamma_t$. The next proposition establishes an equivalent condition for $f\in\Cin(M)$ to be a conserved quantity and the proof can be found in \lco \cite{lco1}:

\begin{proposition}
    A function $f\in\Cin(M)$ is a conserved quantity if and only if it satisfies $\{f,h\} =\{f, h_i\} = 0$ for every $i = 1, \cdots, k$.
\end{proposition}

\section{Kepler Problem with Perturbed Angular Momentum}\label{stochastickepler}

Let $(B_t^1, B_t^2, B_t^3)$ be a Brownian motion in $\R^3$. Let $h: T^*(\R^3\setminus \{0\}) \rightarrow \R$ denote the Hamiltonian of the Kepler problem \[h = \frac{||\pvec||^2}{2} - \frac{1}{||\qvec||}\] and \[\Jvec = (J^1, J^2, J^3) = (q^2p_3 - q^3p_2,q^3p_1 - q^1p_3,q^1p_2 - q^2p_1)\] denote the angular momentum vector. We also assume positive constants $\nu_1, \nu_2$ and $\nu_3$ to take into account the intensity of perturbations along different angular momentum components. We consider the stochastic Hamiltonian system:

\begin{align}\label{equationofmotion3d}
    \del  \Gamma_t &= X_{h}(\Gamma_t) dt + \sum_{i = 1}^3X_{\nu_iJ^i}(\Gamma_t) \del  B_t^i\nonumber\\
    &= X_{h}(\Gamma_t) dt + \sum_{i = 1}^3X_{J^i}(\Gamma_t) \del ( \nu_iB_t^i).
\end{align}
Since $h$ commutes with the components of the angular momentum, we note that $h$ is a conserved quantity for this problem. Let $\Bvec_t$ denote the process $(\nu_1B^1_t, \nu_2B^2_t, \nu_3B^3_t)$ in $\R^3$. Let $\epsilon_{ijk}$ denote the Levi-Civita tensor. Suppose $\Zvec = (Z^1, Z^2, Z^3)$ satisfies $\{Z^i, J^j\} = \sum_{k = 1}^3\epsilon_{ijk} Z^k$ and $\{Z^i,h\} = 0$ for all $i,j\in \{1,2,3\}$. Using Equation \eqref{stratrepoffbrown}
\begin{align*}
    Z^1(\Gamma_t) - Z^1(\Gamma_0) &= \nu_2\int_0^t \{Z^1, J^2\}(\Gamma_t)\del  B^2_t + \nu_3\int_0^t \{Z^1, J^3\}(\Gamma_t)\del  dB^3_t\\
    &= \nu_2\int_0^t Z^3(\Gamma_t)\del  B^2_t - \nu_3\int_0^t Z^2(\Gamma_t)\del  B^3_t
\end{align*}
which implies
\begin{align*}
    \del  Z^1(\Gamma_t) &= \nu_2 Z^3(\Gamma_t)\del  B^2_t - \nu_3Z^2(\Gamma_t)\del  B^3_t\,.
\end{align*}
Using a similar calculation for $v^2(\Gamma_t)$ and $v^3(\Gamma_t)$, we obtain
\begin{align*}
    \del  \Zvec(\Gamma_t) = -\Zvec(\Gamma_t)\times\del  \Bvec.
\end{align*}
Therefore
\begin{equation}\label{normconservation}
    \del ||\Zvec(\Gamma_t)||^2 = 2\Zvec(\Gamma_t)\cdot\del \Zvec(\Gamma_t) = 0\,,
\end{equation}
which shows that $||\Zvec||$ is conserved along solutions and hence $\Zvec(\Gamma_t)$ lies on a sphere. Note that $\Zvec$ is not conserved in general. Letting $\Zvec = \Jvec$ and $\Zvec = \Avec$, where $\Avec = \pvec\times\Jvec - \frac{\qvec}{||\qvec||}$ is the Runge-Lenz vector, we see that $||\Jvec||$ and $||\Avec||$ are conserved along solutions. 

\medskip

The Stratonovich equations for $\qvec(\Gamma_t)$ and $\pvec(\Gamma_t)$ can also be computed from Equation \eqref{stratrepoffbrown}. They are given by
\begin{align}
    \del\qvec(\Gamma_t) &= \pvec(\Gamma_t)dt - \qvec(\Gamma_t)\times \del \Bvec_t\,,\label{equationforposition}\\
    \del\pvec(\Gamma_t) &= -\frac{1}{||\qvec(\Gamma_t)||^3}\qvec(\Gamma_t)dt - \pvec(\Gamma_t)\times\del \Bvec_t\,.
    \label{equationformomentum}
\end{align}
Taking $f = ||q||$ in Equation \eqref{stratrepoffbrown}, we obtain,
\begin{align*}
    ||\qvec||(\Gamma_t) - ||\qvec||(\Gamma_0) &= \int_0^t \frac{(\qvec\cdot\pvec)}{||\qvec||}(\Gamma_s)ds\,.
\end{align*}
By the Fundamental Theorem of Calculus, the left side of this equation is differentiable almost surely and we have,
\begin{equation}\label{equationfornormofgamma}
    \der{t}{||\qvec||(\Gamma_t)} = \frac{(\qvec\cdot\pvec)}{||\qvec||}(\Gamma_t)\,.
\end{equation}
On the other hand, using that $\qvec\cdot (\pvec\times \del \Bvec_t) = -\pvec\cdot(\qvec\times\del\Bvec_t)$, we have
\begin{align*}
    \del(\qvec\cdot\pvec)(\Gamma_t) &= \pvec(\Gamma_t)\cdot\del\qvec(\Gamma_t) + \del\pvec(\Gamma_t)\cdot\qvec(\Gamma_t)\\
    &= \left(||\pvec||^2 - \frac{1}{||\qvec||}\right)dt
\end{align*}which shows that $(\qvec\cdot\pvec)$ is differentiable almost surely. This means that the right side of Equation \eqref{equationfornormofgamma} is differentiable. Using Lagrange's identity $||\Jvec||^2 = ||\qvec||^2||\pvec||^2 - (\qvec\cdot\pvec)^2$ and setting $||\Jvec|| = J = \mathrm{constant}$, we get
\begin{align}\label{deterministicdistance}
    \derk{t}{||\qvec||(\Gamma_t)}{2} &= -\left(\frac{1}{||\qvec||^3}(\qvec\cdot\pvec)^2\right)(\Gamma_t) + \frac{1}{||\qvec||}\left(||\pvec||^2 - \frac{1}{||\qvec||}\right)(\Gamma_t)\nonumber\\
    &= \frac{J^2}{||\qvec||^3(\Gamma_t)} - \frac{1}{||\qvec||^2(\Gamma_t)}.
\end{align}
A similar calculation with the deterministic Kepler problem $\dot{\gamma}(t) = X_{h}(\gamma(t))$ shows that $||\qvec||(\gamma(t))$ also satisfies Equation \eqref{deterministicdistance}. 
Consequently, for almost every $\omega\in\mathcal{S}$, 
\begin{align}
\||\qvec||(\Gamma_t)(\omega) = ||\qvec||(\gamma(t))\,.
\end{align} 
In conclusion, we obtain that $||\qvec||(\Gamma)$ behaves deterministically. Also, since $h$ is conserved along solutions we conclude that $||\pvec||(\Gamma)$  behaves deterministically as well.

\medskip

\subsection{Hamilton’s velocity vector circle  theorem}

Consider now
\[
\Xvec := \frac{\Avec\times\Jvec}{||\Jvec||^2} \,.
 \]
  In case of the deterministic Kepler problem, by direct computation we can show that
   \[\der{t}{||\pvec(\gamma(t)) - \Xvec(\gamma(t))||} = 0\,.\] 
  along any solution $\gamma(t)$ and so $\Xvec$ is the center of the momentum hodographs (see Milnor \cite{milnor}). In case of our stochastic Kepler problem we have
\begin{align*}
    \del \Xvec(\Gamma_t)  &= \frac{1}{J^2}\left(\del \Avec(\Gamma_t)\times\Jvec(\Gamma_t) + \Avec(\Gamma_t)\times\del\Jvec(\Gamma_t) \right)\\
    &=-\frac{1}{J^2}\left((\Avec(\Gamma_t)\times\del\Bvec_t)\times\Jvec(\Gamma_t) + \Avec(\Gamma_t)\times(\Jvec(\Gamma_t)\times\del\Bvec_t)\right)\\
    &= -\frac{1}{J^2}\left((\Avec(\Gamma_t)\times\del\Bvec_t)\times\Jvec(\Gamma_t) - (\Jvec(\Gamma_t)\times\del\Bvec_t)\times\Avec(\Gamma_t) \right)\\
    &= -\frac{1}{J^2}\left((\Avec(\Gamma_t)\times\del\Bvec_t)\times\Jvec(\Gamma_t) + (\del\Bvec_t\times \Avec(\Gamma_t))\times\Jvec(\Gamma_t)+(\Avec(\Gamma_t)\times \Jvec(\Gamma_t))\times\del \Bvec_t\right)\\
    &= -\Xvec(\Gamma_t)\times\del \Bvec_t.
\end{align*}
Hence $||\Xvec||(\Gamma_t)$ is constant and $\Xvec(\Gamma_t)$ evolves on a sphere. Next we calculate
\begin{align*}
    \qvec\cdot\Xvec &= \frac{1}{||\Jvec||^2}[\qvec\cdot (\Avec\times\Jvec)]\\
    &= -\frac{1}{||\Jvec||^2}[\Avec\cdot(\qvec\times\Jvec)]\\
    &= \frac{1}{||\Jvec||^2}[\Avec\cdot(||\qvec||^2\pvec - (\qvec\cdot\pvec)\qvec)]\\
    &= \frac{1}{||\Jvec||^2}[||\qvec||^2(\Avec\cdot\pvec) - (\Avec\cdot\qvec)(\qvec\cdot\pvec)]\\
    &= \frac{1}{||\Jvec||^2}[-||\qvec||(\qvec\cdot\pvec) - ||\Jvec||^2(\qvec\cdot\pvec) +||\qvec||(\qvec\cdot\pvec) ]\\
    &= \qvec\cdot\pvec.
\end{align*}
This implies
\begin{align*}
    \del ||\pvec(\Gamma_t) - \Xvec(\Gamma_t)||^2 &= \del ||\pvec||^2(\Gamma_t) - 2\Xvec(\Gamma_t)\cdot\del\pvec(\Gamma_t) - 2\pvec(\Gamma_t)\cdot\del\Xvec(\Gamma_t)\\
    &= 2\frac{(\qvec\cdot\pvec)(\Gamma_t)}{||\qvec||^3(\Gamma_t)}dt + 2\Xvec(\Gamma_t)\cdot\frac{\qvec}{||\qvec||}(\Gamma_t) \\&+ 2\Xvec(\Gamma_t)\cdot(\pvec(\Gamma_t)\times\del\Bvec_t) + 2\pvec(\Gamma_t)\cdot(\Xvec\times\del\Bvec_t)\\
    &= \frac{2}{||\qvec||^3(\Gamma_t)}[(\qvec\cdot\pvec)(\Gamma_t) + (\Xvec\cdot\qvec)(\Gamma_t)]dt\\
    &= 0.
\end{align*}
Consequently, the momentum vector $\pvec(\Gamma)$ is always a constant distance away from a stochastic process on a sphere. This serves as the analogue of the deterministic Hamilton's theorem (see Milnor \cite{milnor}) for the stochastic Kepler problem. 

\section{Collisions and the Moser Regularization}\label{moserregularization}

In the previous section we showed that if $\Gamma_t$ is a solution of \eqref{equationofmotion3d} then $||\qvec||(\Gamma_t)$ behaves exactly as in the deterministic case. Therefore, akin to the deterministic case, we  define a collision solution as one for which $||\qvec||_t:=||\qvec||(\Gamma_t) \rightarrow 0$ in finite time. Any such solution must have $||\Jvec|| = 0$ throughout since $||\Jvec||$ is conserved. This, in turn, implies that $\Jvec = 0$ along the solution. The converse is true as well, since any solution with $||\Jvec|| = 0$, or equivalently, $\Jvec = 0$ must behave exactly the same as the deterministic Kepler problem by replacing in $J^i = 0$ in equation \eqref{equationofmotion3d}. This means that $||\qvec||_t \rightarrow 0$ in finite time as in the deterministic Kepler problem. In summary, \textit{the solutions ending up in a collision are precisely the ones with angular momentum equal to zero.}

\medskip
This allows us to use the Moser map to regularize the collisions. The procedure is similar to the deterministic Kepler problem, for which we refer to Cushman and Bates \cite{cb} and Heckmann and de Laat \cite{heckdelaat}. We define \[F(\qvec,\pvec):T^*(\R^3\setminus\{0\})\rightarrow\R\] to be the Hamiltonian \[F(\qvec,\pvec) = \frac{1}{2}(||\qvec||) \left(||\pvec||^2+1\right).\] Let $(\Qvec(t), \Pvec(t))$ denotes a solution of the deterministic Kepler problem and define a new time parameter $s$ by $\frac{\mathrm{d}s}{\mathrm{d}t} = \frac{1}{||\Qvec(t)||}$. Then the integral curves of $X_F$ on $F^{-1}(1)$ with respect to the time parameter $s$ are the integral curves of $X_{h}$ on $h^{-1}(-\frac{1}{2})$ with respect to time $t$.

\medskip
Next, let $\Gamma_t$ denote a solution to the stochastic Kepler problem. Consider the stochastic process $Y_t := \left(\int_{0}^t\frac{dt}{||\qvec||_t}, B^1_t, B^2_t, B^3_t\right)$. Since $||\qvec||_t = ||\Qvec(t)||$, we will write the first component of $Y_t$ as $s(t)$ since it is identical to the time parameter $s$ in the previous paragraph. Since on $h^{-1}(-\frac{1}{2}) = F^{-1}(1)$, we have $X_{h}(\Gamma_t) ||\qvec||_t = X_F(\Gamma_t)$, it follows that the equation
\begin{equation}\label{moser1}
    \del \tilde{\Gamma}_t = X_F(\tilde\Gamma_t)ds(t)+ \sum_{i = 1}^3X_{J^i}(\tilde\Gamma_t) \del ( \nu_iB_t^i).
\end{equation}
has the same solutions on $F^{-1}(1)$ as \eqref{equationofmotion3d} on $h^{-1}(-\frac{1}{2})$. It is important to note that the ``true" time parameter in this equation is $t$ and we are not considering the noise to be parametrized by $s$. 

\medskip
Define $K:\kepphase\rightarrow\R$ by $K(\qvec,\pvec) = \frac{1}{2}F^2(\qvec,\pvec)$. Then, on $K^{-1}\left(\frac{1}{2}\right) = F^{-1}(1)$, $X_K$ and $X_F$ are equal, and consequently, solutions to \eqref{moser1} also satisfy
\begin{equation}\label{moser2}
    \del \tilde{\Gamma}_t = X_K(\tilde\Gamma_t)ds(t)+ \sum_{i = 1}^3X_{J^i}(\tilde\Gamma_t) \del ( \nu_iB_t^i).
\end{equation}
The Moser map is the map \[\phi_M:\kepphase\rightarrow\{(\mathbf{u}, \mathbf{v})\in T^*{S^3}|\:\:\uvec\neq(0,0,0,1), \vvec\neq 0\} =: \TxS\]given by 
\begin{align*}\phi_M(\qvec,\pvec) = \left(\left(\frac{2\pvec}{||\pvec||^2+1}, \frac{||\pvec||^2-1}{||\pvec||^2+1}\right),\left(-(||\pvec||^2+1)\frac{\qvec}{2}+(\qvec\cdot\pvec)\pvec,-\qvec\cdot\pvec\right)\right).\end{align*}
The Moser map is a symplectomorphism, and in particular, restricts to a symplectomorphism between $h\inv(-\frac{1}{2})$ and $\ToneS:= \{(\uvec,\vvec)\in\TxS|\:\:||\vvec||^2 = 1\}$. On $\ToneS$, consider the Hamiltonian obtained by pushforwarding $K$ on $K^{-1}\left(\frac{1}{2}\right)=h^{-1}(-\frac{1}{2})$ by $\phi_M$, namely, we let \[G(\uvec,\vvec):= K\circ\phi_M^{-1} = \frac{1}{2}||\vvec||^2.\]$G$ can be extended to a smooth function on $T^*S^3\bigcap G^{-1}(\frac{1}{2})$, which we also denote by $G$. 

\medskip
The Hamiltonian $G$ is symmetric under the cotangent-lifted $SO(4)$ action on $G\inv\left({\frac{1}{2}}\right)$. In particular, each of the components of \[\lvec(\uvec, \vvec) = (\lambda^1(\uvec, \vvec), \lambda^2(\uvec, \vvec), \lambda^3(\uvec, \vvec)) := (u^2v_3 - u^3v_2, -u^1v_3 + u^3v_1, u^1v_2 - u^2v_1)\] are conserved. The Moser map pulls back $\lambda^i$ to the angular momentum components $J^i$, for each $i = 1,2,3$, restricted to $h^{-1}(-\frac{1}{2})$. 

\medskip
We now turn our attention back to the stochastic Kepler problem \eqref{equationofmotion3d}. Using the Moser map, and noting the fact that the Moser map pushes forward the components of $J^i$ to $\lambda^i$, we obtain the following stochastic perturbation of the geodesic flow on $G\inv(\frac{1}{2})$:
\begin{equation}\label{moser3}
    \del \tilde{\Gamma}_t = X_G(\tilde\Gamma_t)ds(t)+ \sum_{i = 1}^3X_{\lambda^i}(\tilde\Gamma_t) \del ( \nu_iB_t^i).
\end{equation}
Note that $\lambda^i$ is not a conserved quantity for Equation \eqref{moser3}. However, arguing similar to $||\Jvec||$ as in the previous section, we can show that $||\lvec||$ is conserved for Equation \eqref{moser3}. The next theorem shows that the collision solutions of the stochastic Kepler problem with $h = -\frac{1}{2}$ map to great circles passing through the north pole $(0,0,0,1)$ of $S^3$, exactly as in Moser regularization of the deterministic Kepler problem.

\medskip

\begin{theorem}
    A solution $\tilde\Gamma_t$ of \eqref{moser3} has $||\lvec||(\tilde\Gamma_t) = 0$ if and only if it is a collision solution of the stochastic Kepler problem \eqref{equationofmotion3d}. 

    \medskip
    Moreover, the set $||\lvec||\inv(0)$ is the set of all solutions of equation \eqref{moser3} that pass through the set \[C:= \{(\uvec,\vvec)\in\TxS|\:\: \uvec = (0,0,0,1)\}.\] If $\tilde\Gamma_t$ is such a solution, then the projection of $\tilde\Gamma_t$ on $S^3$ is a great circle passing through $(0,0,0,1)$. 
\end{theorem}

\textbf{Proof}: Let $\tilde\Gamma_t$ be a solution of $\eqref{equationofmotion3d}$ on $G^{-1}(\frac{1}{2})$. Since the Moser map pulls back the components of $\lvec$ to the corresponding components of $\Jvec$, it follow that has $\lvec(\tilde\Gamma_t) = 0$ if and only if the corresponding process $\Gamma_t$ obtained by pulling back $\tilde\Gamma_t$ by the Moser map is a solution of the stochastic Kepler problem \eqref{equationofmotion3d} on $h^{-1}(-\frac{1}{2})$ with $\Jvec(\Gamma_t) = 0$. This is possible if and only if $\Gamma_t$ is a collision solution.

\medskip

First, note that $C$ is a subset of $||\lvec||\inv(0)$ since the first three components of $\uvec$ are 0. So we need to prove the reverse inclusion. If $||\lvec||(\tilde\Gamma_t) = 0$, then $\tilde\Gamma_t$ satisfies 
\begin{equation*}
    \del  \tilde\Gamma_t = X_G(\tilde\Gamma_t)ds(t).
\end{equation*}
Recalling $s(t) = \frac{dt}{||\qvec||_t}$ and recalling that $||\qvec||_t$ is deterministic and differentiable, the previous equation is the following ODE:
\begin{equation}\label{moser4a}
    \frac{d\gamma}{dt} = X_G(\gamma(t))\frac{1}{||\qvec||_t}.
\end{equation}
Now we can reintroduce the time parameter $s$ given by $\frac{ds}{dt} = \frac{1}{||\qvec||_t}$. This yields,
\begin{equation}\label{moser4}
    \frac{d\gamma}{ds} = X_G(\gamma(s)).
\end{equation}
 The remainder of the proof proceeds similarly to the proof of (4.10) in Cushman and Bates \cite{cb}. The projection of $\gamma(s)$ on $S^3$ is a great circle. Such a great circle always intersects the equator of $S^3$ given by \[\{\uvec = (u^1, u^2, u^3, u^4)\in S^3|\:\: u^4 = 0\}.\] Let $(\tilde{\uvec}, 0, \tilde\vvec, v_4)$ be the point of intersection. Since $\lvec = 0$ and $\tilde{\uvec}\neq0$, we have $\tilde\vvec = 0$ or $\tilde\uvec = \mu\tilde\vvec$ for some non-zero $\mu\in\R$. The latter implies that $(\tilde\uvec,0)\cdot(\tilde\vvec, v_4) = \mu$, and since $(\uvec,\vvec)\in T^*S^3$, it follows that $\mu = 0$. But then $\tilde\uvec = 0$, which is impossible. Hence, $\tilde\vvec = 0$. Moreover, since equation \eqref{moser3} is defined on $G\inv\left({\frac{1}{2}}\right)$, we obtain $v_4^2 = 1$, or $v_4 = \pm 1$. 

\medskip

Now, we solve the ODE \eqref{moser4} with initial conditions $(\tilde\uvec, 0, 0, v_4)$. Keeping in mind that $G = \frac{1}{2}$, we have \[\gamma(s) = (\cos(s)\tilde\uvec, \sin(s)v_4, -\sin(s)\tilde{\uvec}, \cos(s)v_4)\] At time $s = \pi - v_4\frac{\pi}{2}$, $\cos(s) = 0$ and $\sin(s) = v_4$, so $\gamma(s)$ reaches $C$. This completes the proof. 

\bigskip
This procedure can be extended to any fixed energy level $k<0$ by rescaling $(\qvec,\pvec)\in\kepphase$ and the noise parameters $\nu_1$, $\nu_2$ and $\nu_3$. Indeed, given $a\in(0, \infty)$, rescale time by $t\mapsto a^3 t$, $(\qvec, \pvec)$ by $(\qvec, \pvec) \mapsto \left(a^2\qvec, \frac{1}{a}\pvec\right)$ and the noise intensities $\nu_1, \nu_2, \nu_3$ by $\nu_i \mapsto \tilde\nu_i = \nu_ia^{-\frac{3}{2}}$. The transformation of the phase space variables yields a symplectomorphism $F:\kepphase\rightarrow\kepphase$. The usual symplectic form $\Omega$ on $\kepphase$ is pushed forward by $F$ to $\tilde\Omega= F_*\Omega = a\Omega$. Moreover, $h$ induces a Hamiltonian $\tilde h = h\circ F^{-1}$ on $(\kepphase, \tilde\Omega)$. We have, $\tilde h \left(\qvec, \pvec\right) = a^2h(\qvec, \pvec)$. Similarly, defining $\tilde \Jvec = \Jvec\circ F^{-1}$, we obtain $\tilde \Jvec(\qvec, \pvec) = \frac{1}{a}\Jvec(\qvec, \pvec)$. This implies that the Hamiltonian vector fields of $h$ on $(\kepphase, \Omega)$, denoted by $X^{\Omega}_h$, and $\tilde h$ on $(\kepphase, \tilde\Omega)$, denoted by $X^{\tilde\Omega}_{\tilde h}$ are related by $X^{\tilde\Omega}_{\tilde h} = \frac{1}{a^3}X^{\Omega}_h$. Similarly, we have, $X^{\tilde\Omega}_{\tilde {J}^i} = X^{\Omega}_{J^i}$. Next, using the scaling property of Brownian motion, that is, for any positve constant $c$, $B_{ct} = \sqrt{c}B_t$, we find that $B^i_{a^3t} = a^{\frac{3}{2}}B^i_t$. Therefore, letting $\tilde t = a^3t$, the solutions of 
\begin{equation}\label{moser5}
    \del  \Gamma_{\tilde t} = X^{\tilde\Omega}_{\tilde h}(\Gamma_{\tilde t}) d{\tilde t} + \sum_{i = 1}^3X^{\tilde\Omega}_{\tilde J^i}(\Gamma_{\tilde t}) \del ( \tilde\nu_iB_{\tilde t}^i).
\end{equation}
are identical to those of equation \eqref{equationofmotion3d}. We can then choose $a = \sqrt{-2k}$ to transform the stochastic Kepler problem with energy $k$ to the stochastic Kepler problem with energy $-\frac{1}{2}$.

\bibliographystyle{unsrt}
\bibliography{refer}
\end{document}